\documentclass[reprint,amsmath,amssymb,floatfix]{revtex4-1}
\usepackage{graphicx}
\usepackage{dcolumn}
\usepackage{bm}

\begin{document}
\title{Phase retrieval with complexity guidance}


\author{Mansi Butola}
\author{Sunaina}
\author{Kedar Khare} 
\email{kedark@physics.iitd.ac.in}
\affiliation{Department of Physics, Indian Institute of Technology Delhi, New Delhi 110016 India}

\begin{abstract}
Iterative phase retrieval methods based on the Gerchberg-Saxton (GS) or Fienup algorithm require a large number of iterations to converge to a meaningful solution. For complex-valued or phase objects, these approaches also suffer from stagnation problems where the solution does not change much from iteration to iteration but the resultant solution shows artifacts such as presence of a twin. We introduce a complexity parameter $\zeta$ that can be computed directly from the Fourier magnitude data and provides a measure of fluctuations in the desired phase retrieval solution. It is observed that when initiated with a uniformly random phase map, the complexity of the Fienup solution containing stagnation artifacts stabilizes at a numerical value that is much higher than $\zeta$. We propose a modified Fienup algorithm that uses a controlled sparsity enhancing step such that in every iteration the complexity of the resulting solution is explicitly made close to $\zeta$. This approach which we refer to as complexity guided phase retrieval (CGPR) is seen to significantly reduce the number of phase retrieval iterations required for convergence to a meaningful solution and automatically addresses the stagnation problems. The CGPR methodology can enable new applications of iterative phase retrieval that are considered practically difficult due to large number of iterations required for a reliable phase recovery.
\end{abstract}

\maketitle

\section{Introduction}
Phase retrieval problem arises in different applications like coherent X-ray diffraction imaging \cite{Nugent, Chapman}, astronomical imaging \cite{Dainty}, lens-less imaging \cite{Fienuplensless} , microscopy \cite{Ferraro}, optical encryption \cite{Johnson}, beam shaping \cite{Marozas} and optical communication \cite{Ivankovski}, etc. Interferometry is one way to recover the phase by superimposing the unknown object beam with a reference beam, however, this modality requires sophisticated experimental set-up that may not always be possible to build for wavelengths beyond visible light. Another way to recover the phase information is through iterative phase retrieval algorithms that deal with the problem of recovery of a complex-valued object from its Fourier magnitude. The idea of phase recovery from the Fourier intensity was enunciated by Sayre while studying the Bragg diffraction peaks of the crystal structures \cite{Sayre}. Gerchberg and Saxton \cite{Gerchberg} developed an early algorithm to reconstruct the object from the Fourier modulus along with specified object constraints. Fienup \cite{Fienup1, Fienuptour} modified the GS algorithm by adding a negative feedback term in the update and showing significant improvement in its convergence rate. This algorithm is famously known as the hybrid input-output (HIO) algorithm. A generalization of the Fienup algorithm was presented in the form of difference maps \cite{Elser}. Recent developments in the the field of phase retrieval include the non-linear sparsity based algorithms like GESPAR \cite{GESPAR}, PhaseLift \cite{PhaseLift}, Wirtinger flow \cite{candes}, etc. that have been reviewed in \cite{eldar}. 

The iterative phase retrieval algorithms usually operate by imposing measured Fourier intensity constraint in the data domain and some other suitable constraints such as object support, positivity, sparsity, etc. in the image domain. However it may be observed that the constraints in the data and image domains do not have any connect. Additionally it is well-known empirically that the GS or Fienup algorithms and their variants require large number of iterations to converge to a desirable solution when initiated with uniform random phase map. In X-ray coherent diffractive imaging (CDI) used in imaging of non-crystalline and nano structures \cite{Miao}, the standard procedure starts with multiple initial random guesses for phase solution followed by generations of phase retrieval iterations amounting to over hundred thousand of iterations involving forward and backward Fourier transforming operations. Considering the importance of phase retrieval for a number of topical areas, the requirement of large number of phase retrieval iterations can be a roadblock in designing associated applications or devices.

The GS and Fienup algorithms are known to suffer from twin-image stagnation problem when the desired solution is complex-valued. This problem arises since a function  $g(x,y)$ and its twin $g^*(-x,-y)$ both have the same Fourier magnitude $|G(f_x,f_y)|$. Stagnation is troublesome since the solution makes little progress from iteration to iteration suggesting ``convergence'' of the algorithm but the resultant solution is not necessarily meaningful. In order to address the issue of twin-image stagnation some approaches have been reported in the literature. By fabricating a non-centrosymmetric object support or by making a diversity of measurement one can significantly reduce the chances of twin-image stagnation\cite{Wackerman, Guizar}. Movable aperture lensless transmission microscopy \cite{Faulkner} can retrieve the phase without twin-image stagnation by taking a series of diffraction patterns which can be done by scanning an aperture over the object wave function to two or more positions. A phase perturbation method was suggested in \cite{sheridan} for avoiding stagnation problems.  A sparsity-assisted phase retrieval algorithm \cite{Gaur} has also been proposed to address the twin image stagnation issue.

In the present study we observe the nature of the phase retrieval solution as the iterations progress in terms of what we call as the complexity of the solution. The complexity as we will define in the next section, is a numerical parameter that is a measure of the fluctuations in the solution. We show that the complexity of the desired solution can be measured \textit{a priori} from the Fourier magnitude data. This complexity information which is available from the Fourier magnitude data has not been used in the phase retrieval problem to the best of our knowledge. In a sparsity assisted phase retrieval framework, we show that the complexity parameter can be used explicitly as a guidance to control the solution sparsity. This methodology is seen to significantly reduce the number of phase retrieval iterations required to reach a meaningful solution and is additionally seen to address the twin-stagnation problem.

This paper is organized as follows. In Section 2, we define the complexity parameter and a method to compute it from the Fourier magnitude data. In Section 3 we study the behavior of the solution as a function of iterations in terms of the complexity parameter. Section 4 presents a step-by-step description of complexity guided phase retrieval(CGPR) algorithm that uses the complexity parameter explicitly for guiding the sparsity enhancement step in the iterations. A comparative study of the Fienup HIO and CGPR is shown for a binary phase object. In Section 5, the performance of CGPR is shown for the reconstruction of binary phase object for different noise levels in Fourier intensity data. Finally in Section 6, we briefly summarize our findings regarding phase retrieval algorithm based on complexity guidance and comment on practical applications of this methodology.  
 
\section{Estimating solution complexity from Fourier magnitude data}
\label{sec:examples}
We begin this section by defining a complexity parameter a complex-valued object $g(x,y)$, that provides a measure of fluctuations in its pixel values. The complexity $\zeta$ for $g(x,y)$ may be defined as:
\begin{equation}\label{complexity_def}
\zeta = \sum_{i=all pixels} (| \nabla _x g_i |^2 + | \nabla _y g_i |^2),
\end{equation} 
where, $\nabla _x$ and $\nabla _y$ are the $x$ and $y$ gradient operators respectively that may be implemented numerically by central differencing scheme. In order to calculate the complexity of an object $g(x,y)$, the object itself should be known. However, in problems like phase retrieval, we have access to the Fourier intensity $ |G(f_x,f_y)|^2 $ of the object and not the object $g(x,y)$. Interestingly in this case, complexity of the object $g(x,y)$ can still be calculated using the Fourier magnitude information. The derivative property  of Fourier transform when applied along with the Parseval's theorem can give us the numerical value of $\zeta$ as per the following relation:
\begin{align}\label{eq:wrong_id}
\int \int  (| \nabla _x g | ^2 &+ | \nabla _y g | ^2 ) dx dy \\ 
                               &= \int \int ( |i 2\pi f_x G |^2 + |i 2\pi f_y G|^2 ) d f_x d f_y.
\end{align}
In a practical phase retrieval setup, the measured Fourier magnitude $|G(f_x, f_y)|$ as well as the recovered solution $g(x,y)$ are both discrete in nature. The complexity $\zeta$ of any given $g(x,y)$ therefore can be evaluated using the central differencing scheme.
\begin{equation}\label{eq:CDF}
\nabla_x g(x,y) = \frac{g(x+\Delta x,y)-g(x-\Delta x,y)}{2 \Delta x},  
\end{equation}
with a similar relation for the $y$-derivative. Taking Fourier transform on the both sides of the above equation and employing shifting property of Fourier transform we get,
\begin{align}
\mathcal{F}(\nabla_x g(x,y)) &= G(f_x,f_y)\frac{ exp(i2\pi f_x\Delta x)- exp(-i2\pi f_x\Delta x)}{2 \Delta x} \\
                             &= i \frac{\sin(2\pi f_x \Delta x)}{\Delta x} G(f_x,f_y)
\end{align}     
To get an equivalent numerical value for $\zeta$ from the Fourier magnitude data, we have to use the 'modified wave number' relation as given below \cite{Rajora}:
\begin{equation}\label{eq:cor_derivative_id}
\zeta = \sum_{i = all pixels} [\frac {\sin^2 (2 \pi f_{xi} \Delta x)}{\Delta x^2} + \frac {\sin^2 (2 \pi f_{yi} \Delta y)}{\Delta y^2}]|G_i|^2 ,
\end{equation}
where $\Delta x$ and $\Delta y$ are the sampling intervals in $x$ and $y$ direction respectively in the image space. The multipliers $\frac{\sin ((2 \pi f_x \Delta x )}{\Delta x} \textrm{and}  \frac{\sin ((2 \pi f_y \Delta y )}{\Delta y} $ are referred as 'modified wave number' and they reduce to $2\pi f_x$ and $2\pi f_y$ respectively as $\Delta x, \Delta y \to 0$. It is easy to verify that the numerical values of $\zeta$computed using Eq. (\ref{complexity_def}) and Eq.(\ref{eq:cor_derivative_id}) are equal. The complexity information captured in the parameter $\zeta$ is typically not used by the standard phase retrieval algorithms, but is very valuable as we will see later in the paper. 

\begin{figure}[htbp]
\centering
\includegraphics[width = 0.50\textwidth]{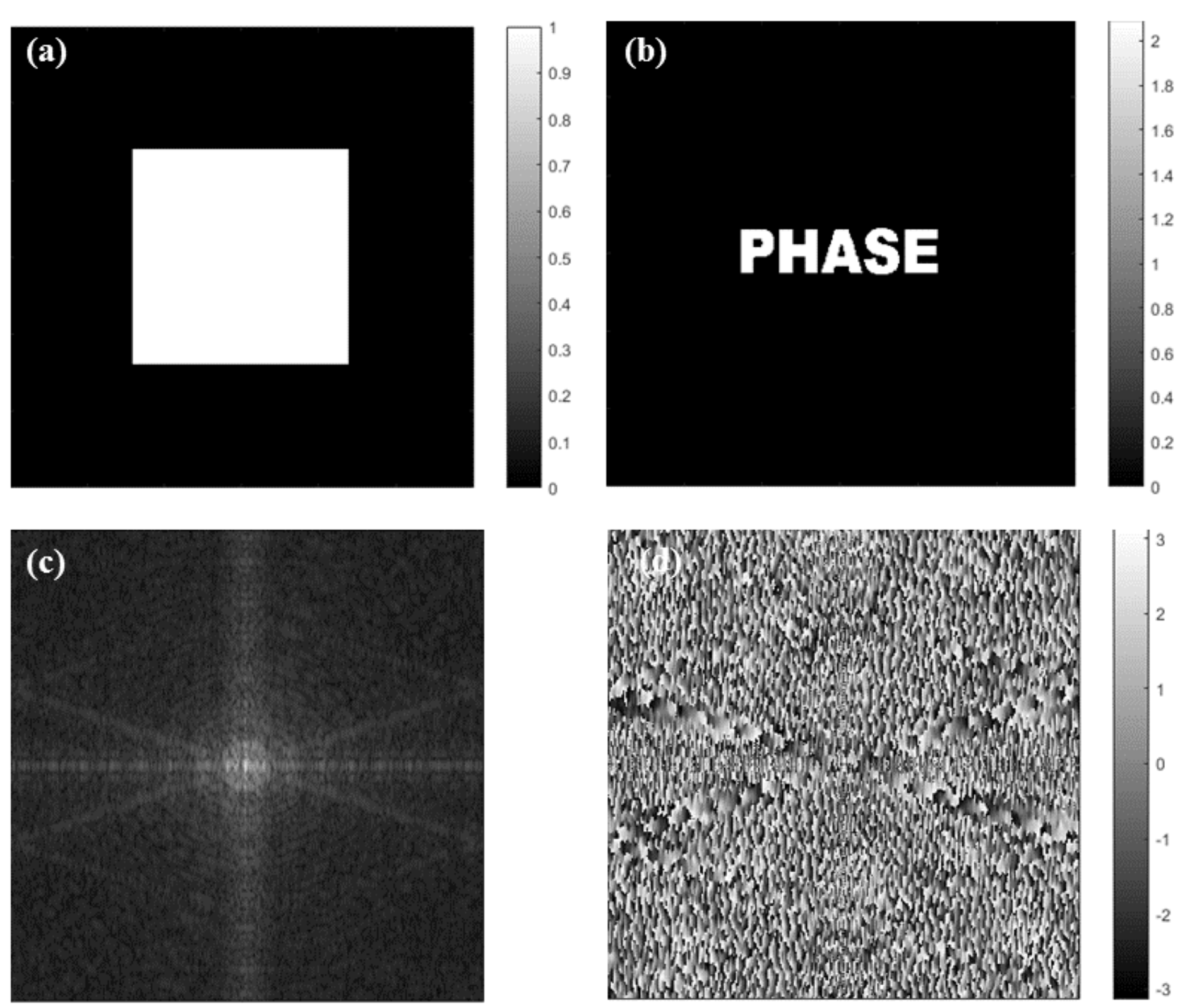}
\caption{{Binary phase object of size $280 \times 280$, with a phase step of  $\frac{2 \pi}{3} $, is defined over computational window of size $600 \times 600$. Amplitude and phase of the object are shown in (a) and (b) respectively. \textbf{(c)} Fourier magnitude $| G |$ and \textbf{(d)} Fourier phase $arg(G)$ corresponding to the object $g(x,y)$}}
\label{phase_object}
\end{figure}

For illustration we take a unit amplitude binary phase object "PHASE" inside a computational window of size $600\times 600$ as shown in Fig. (\ref{phase_object}). The amplitude and phase of the object are shown in Fig. (\ref{phase_object}) (a), (b) respectively. A phase step of $\frac{2 \pi}{3} $ is used. The object support window has an extent of $ 280 \times 280 $ in the center, which is less than half of the computational window size, to ensure that the Fourier intensity is Nyquist sampled. The magnitude and phase of the 2D Fourier transform of the object are shown in Fig. (\ref{phase_object})(c), (d) respectively. The complexity parameter $\zeta$ calculated using Eq. (\ref{complexity_def}) and Eq. (\ref{eq:cor_derivative_id}) is seen to match to double precision. 
\section{Behavior of complexity of solution as a function of phase retrieval iterations}
With the definition of complexity from previous section it is apparent to us that a random-patterned image will have very high fluctuations and thus has high complexity parameter. On contrary, a constant image will have no fluctuations thus has very low complexity parameter. 
In traditional Fienup algorithm, we typically start either with a random phase guess solution having very high complexity or with a constant guess solution having very low complexity. The complexity of the starting guess is therefore completely different from the desired solution complexity that can be estimated from the Fourier magnitude data. For illustration we take the binary phase object as shown in Fig. 1(a), (b) and using the corresponding Fourier magnitude data $|G(f_x,f_y)|$ (as in Fig. 1(c)) observe the complexity of resultant solutions as a function of HIO iterations. Starting with a constant phase function and a uniform random phase function with phase distributed in $[0, 2\pi]$ we ran $500$ HIO iterations each and the corresponding phase solutions after $500$ iterations are shown in Fig. 2(a), (b) respectively. As the HIO iterations progressed, we computed the complexity of the guess solutions using Eq. (\ref{complexity_def}) and the corresponding plots of logarithm of the complexity parameter for the two cases are shown in Fig. 2(c) (blue and red curves). The constant green line in this plot shows the desired complexity value of the solution estimated using the Fourier magnitude data as per Eq. (\ref{eq:cor_derivative_id}). The inset in Fig. 2 shows a zoomed version of the complexity plots for initial $100$ iterations. We see that for both the illustrated cases (constant initial guess and random initial guess), the solution has stabilized by $500$ iterations but has several artifacts including twin stagnation. The complexity parameter for both the solutions has also stabilized to a numerical value which approximately $25 \%$ higher than the desired complexity obtained from the Fourier magnitude data as depicted by the green curve. We make an important observation that the artifacts in the HIO solutions (Fig. 2(a), (b)) can be associated with the higher complexity value for these solutions. Simply continuing with more number of HIO iterations is not likely to reduce the complexity of the corresponding solutions and hence not mitigate the artifacts. The observations made above are generic and they suggest that if the complexity of the solution is brought close to the desired complexity level in each iteration, this may provide a way out of the well-known stagnation issues. In the next section we will add a controlled sparsity enhancement step in the HIO iteration and observe the effect of this on the convergence of phase retrieval solution. 
\begin{figure}[htbp]
	\includegraphics[width =0.50 \textwidth]{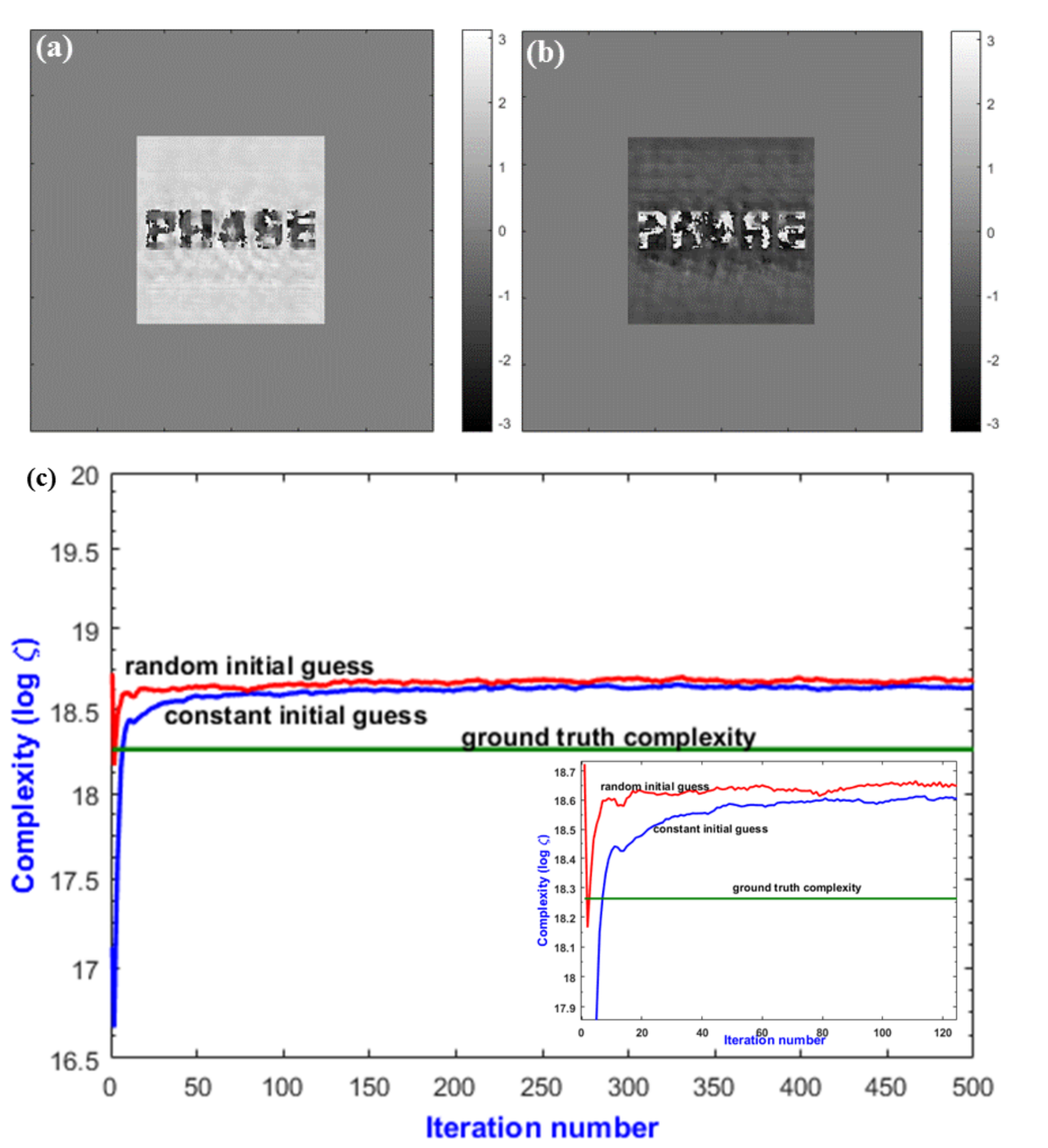}
	\caption{ Behavior of complexity with number of iterations.\textbf{(a)} Reconstruction of binary phase image shown in Fig. 1(a) by  HIO algorithm when an initial constant guess for phase is used and \textbf{(b)} a random initial guess is used. \textbf{(c)} The plot of complexity parameter versus number of iterations for constant and random initial guess shows that these complexities never reach to the complexity of the ideal solution which is represented by the green curve. }
\end{figure}
 
\section{Complexity-guided phase retrieval}
From the previous section, we understand that the complexity of the guess solution never matches with the complexity $\zeta$ even with a very large number of HIO iterations. The complexity parameter thus provides an interesting way of understanding the stagnation issues in phase retrieval algorithms. It has been shown that a sparsity enhancing step added to the traditional HIO algorithm helps eliminate the twin-image problem. In \cite{Gaur} a fixed number of sparsity enhancing sub-iterations were used in each HIO iteration to demonstrate elimination of the twin image. This approach while sufficient for demonstration of twin image elimination still requires a large number of iterations for convergence to a meaningful solution. In this work we present a phase retrieval method with an adaptive sparsity enhancing step guided by the complexity parameter and call it `complexity-guided phase retrieval(CGPR)'. In CGPR, sparsity enhancing steps are added to HIO in a controlled fashion such that in each HIO iteration, the complexity of the solution nearly matches with the complexity parameter $\zeta$ computed as per Eq. (\ref{eq:cor_derivative_id}). We make an interesting observation that CGPR requires significantly reduced number of total iterations for reaching a reasonable solution. 

For the present work where a binary phase object as in Fig. 1 is used for illustration, we use the total variation (TV) of the solution as a sparsity criterion which is widely being used for sparse recovery problems. The TV functional is defined as:
\begin{equation}
TV(g, g*) = \sum_{i=all pixels} \sqrt{| \nabla _x g_i |^2 + | \nabla _y g_i |^2}.
\end{equation}
Since $g$ is complex valued, the functional gradient of TV to be used in a gradient descent scheme for TV reduction is evaluated with respect to the conjugate image $g*$ and is given by \cite{Gaur}:
\begin{equation}
\nabla_{g*} TV(g,g*)= - \frac{1}{2}\nabla . (\frac{\nabla g}{|\nabla g|}). 
\end{equation}
For convenience in the following description, we define a unit vector in the direction of the functional gradient of TV as:
\begin{equation}
\hat{u} = \frac{\nabla_{g*} TV(g,g*)}{|| \nabla_{g*} TV(g,g*) ||_2}.
\end{equation} 

We now discuss the detailed steps involved in CGPR for the reconstruction of a phase object $g(x,y)$. Given the Fourier magnitude $|G(f_x,f_y)|$, firstly the complexity parameter $\zeta$ is calculated as per Eq. (\ref{eq:cor_derivative_id}). An initial guess is made where we start with a phase function $g_0 = exp(i\theta_{0})$, with $\theta_0(x,y)$ being a random phase map having phase values uniformly distributed in $[0, 2\pi]$. 
\begin{enumerate}
\item For $(n+1)th $ iteration, we evaluate the Fourier transform of the previous guess solution  $g_n$ given by$\mathcal{F}{g_n}= \hat{G}_n = |\hat{G}_n| \exp(i2 \pi \hat{\phi}_n (f_x,f_y))$ and replace the Fourier magnitude of $\hat{G}_n$ with the given Fourier magnitude $|G|$, keeping the phase unchanged. 
\begin{equation}
\hat{G}_n = |G| exp (i2 \pi \hat{\phi}_n (f_x,f_y)).
\end{equation}
\item The inverse Fourier transform of $\hat{G}_n$ is computed as $\hat{g}'_n = \mathcal{F}^{-1}\hat{G}_n$.
\item  The resultant solution is updated with the Fienup HIO step with negative feedback:
\begin{align}
\hat{g}_{n+1}  &=\hat{g}'_n& ,  (x, y) \in C \\ 
             &= \hat{g}'_n - \beta \hat{g}_n    &,(x,y) \notin C   
\end{align}
 where C is the support constraint in object domain. The $\beta$ parameter has value in $(0.5,1)$ and here we have taken $\beta = 0.9$.
\item The TV of this intermediate solution is reduced by recursively applying a number of small gradient descent steps to the group of pixels in $\hat{g}_{n+1}$ that belong to the support $C$. The $(k+1)th$ gradient descent step is:
\begin{equation}
\hat{g}_{n+1}^{k+1} = \hat{g}_{n+1}^{k} - t || \hat{g}_{n+1}^k||_2 \: [ \hat{u} ]_{g = \hat{g}_{n+1}^k }.
\end{equation}
We have chosen $t=0.005$ so that the solution progresses by a small amount in each gradient descent step. The TV reduction iteration is started with $\hat{g}_{n+1}^{0} = \hat{g}_{n+1}$. After each gradient descent step in the TV-reducing direction, we compute the complexity parameter $\zeta_{n+1}^{k+1}$ for $\hat{g}_{n+1}^{k+1}$ in the image domain where the image gradients are calculated by central differencing scheme as per Eq. (\ref{complexity_def}). The gradient descent steps for TV reduction are stopped when the complexity parameter $\zeta_{n+1}^{k+1}$ for the updated solution $\hat{g}_{n+1}^{k+1}$ is within $0.5 \%$ of the desired value $\zeta$. In this way, the number of sparsity enhancing steps are controlled by the complexity parameter.
\item At this point the $(n+1)th$ iteration is completed and we set $g_{n+1} = \hat{g}_{n+1}^{k+1}$.
\end{enumerate}
To draw the comparison between HIO and CGPR algorithms the phase recovery results for the binary phase object (shown in Fig. 1(a)) are shown in Fig. 3. Here, both the algorithms have been initiated with the same initial random phase map. Figures 3(a), (b) show the reconstructed phase image with $500$ and $1000$ iterations of the HIO algorithm where only the support constraint has been used. The reconstructed phase after $200$ and $500$ iterations of CGPR is shown in Fig. 3 (c), (d) respectively.  We can see that even with $200$ iterations of CGPR algorithm, the solution in Fig. 3(c) is much better than that in Fig. 3(a), (b) with almost no perceivable stagnation artifacts. Figure 3(e) shows a plot of logarithm of the error as a function of iteration number for the first $500$ iterations of the HIO and CGPR algorithms. 
The error metric is defined in the object domain as \cite{Fienup2}: 
\begin{equation}
E^2  = min(E^2({g}_n(x,y)),E^2({g}_n^*(-x,-y))),
\end{equation}
\begin{equation}
where, E^2({g}_n(x,y)) =  \frac{\sum |{g_n}|^2+ \sum |g|^2 -2 corr({g}_n,g)}{\sum |g|^2}.
\end{equation}
Here ${g}_n$ is the guess solution after $n$ iterations, $g$ is the ground truth object and $corr({g}_n,g)$ is the correlation of ${g}_n$ and $g$. One can clearly see that with CGPR the error falls rapidly to a much smaller value compared to the HIO algorithm. In $20$ trials of CGPR, initiated each time with different random guess, we observed that the object was always recovered very well on an average in $200$ iterations whereas HIO iteration alone could not produce similar quality reconstructions. The computational time taken by the $200$ iterations of HIO and CGPR was observed to be $4$ seconds and $57$ seconds respectively in MATLAB environment (running on a desktop computer with $3.5$ GHz processor with $16$ GB RAM). Though the HIO algorithm takes much less computational time compared to CGPR for the same number of iterations, it generally does not seem to promise a good recovery for phase objects even if the number of iterations is made large. With CGPR we can almost seem to guarantee a reasonable recovery after a fixed number of iterations. In addition to this benefit CGPR is seen to automatically eliminate the twin-image as well.
 \begin{figure}[htbp]
\centering
\includegraphics[width = 0.50 \textwidth]{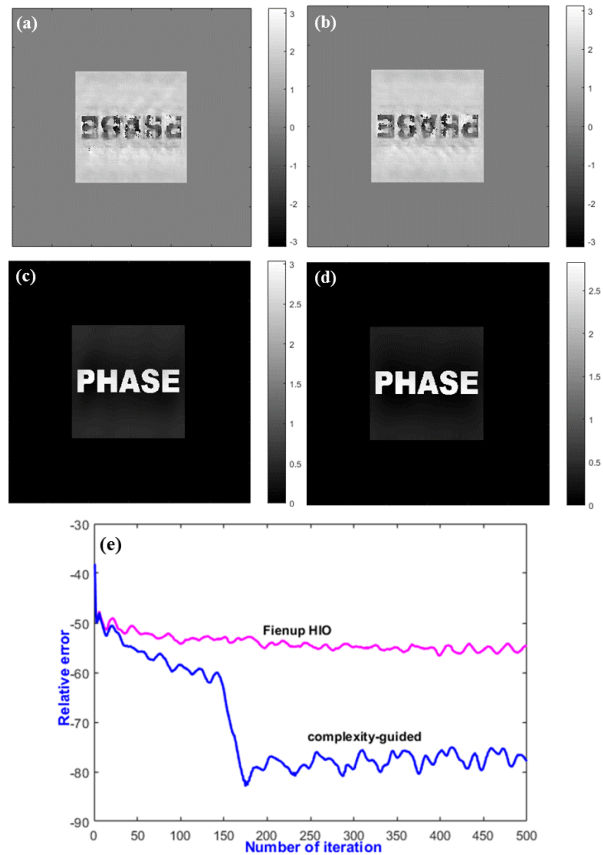}
\caption{Comparison of HIO with complexity guided phase retrieval.{(a)} Reconstruction of binary phase image by HIO algorithm in 500 and {(b)} 1000 iterations. {(c)}The phase recovery by CGPR with 200 iterations and  {(d)} with 500 iterations.  {(e)} 10 times logarithm of error metric plotted against the number of iterations for HIO and CGPR algorithms.  }
\end{figure}

\section{Effect of Noise on Complexity Parameter }
It has now become quite clear that the heart of our algorithm is – applying the sparsity to the guess solution until the complexity parameter of the guess solution matches with that of the object. In section 2, we have discussed how to calculate the complexity parameter $\zeta$ from the Fourier magnitude data. In practice, the far-field diffraction intensity of the object detected on an array sensor will be corrupted by Poisson noise even if other forms of noise - e.g. due to electronic readout process - is minimized. To study the effect of noise on the reconstructed image using CGPR, we have generated Fourier intensity data with Poisson noise corresponding to a light level of $10^4$ and $10^6$ photons/pixel. The numerical values of the complexity parameter $\zeta$ for the normalized Fourier intensity data corresponding to the two light levels are observed to be nearly equal (within $0.01 \%$ of each other). The phase reconstruction results with HIO and CGPR algorithms for the noisy Fourier data generated for two light levels ($10^4$ and $10^6$ photons/pixel) are shown in Fig. 4 (a), (b) and Fig. 4 (c),(d) respectively. We can see that the quality of reconstruction with CGPR is reasonable for both the noise levels and is much better than what is obtained with HIO alone. For lower light levels we observe that due to the strong dc peak in the Fourier intensity, high spatial frequency part of the data is essentially noise. Any phase retrieval scheme which is based on directly replacing the Fourier magnitude with measured data in Fourier space may not be the best in such case. For Fourier intensity data having significant Poisson noise, alternative modeling approaches based on optimization ideas may be suitable. But even in such cases, the complexity guidance idea may prove to be useful. We will explore this direction in future.  
\begin{figure}[htbp]
\centering
\includegraphics[width = 0.50 \textwidth]{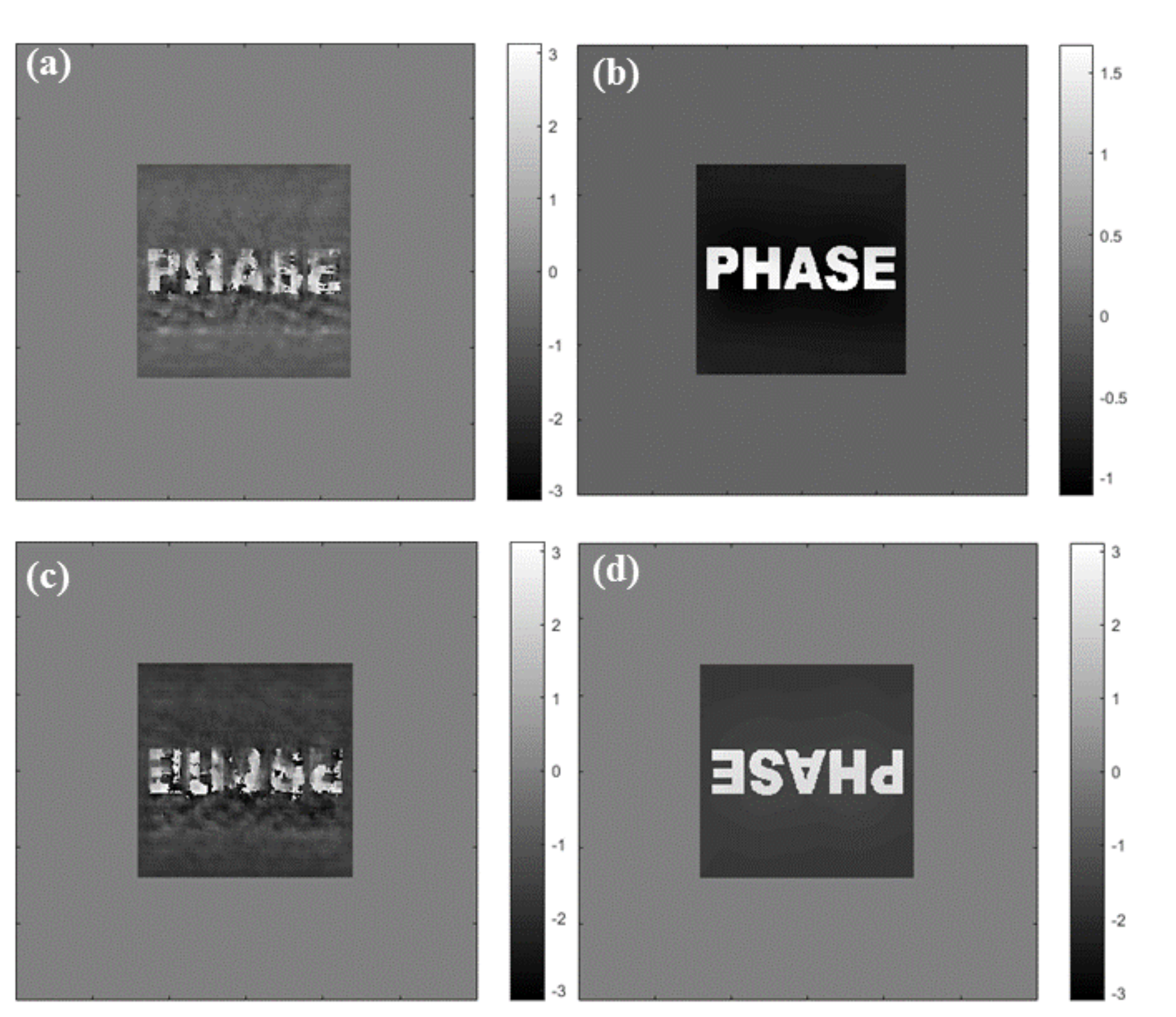}
\caption{Phase reconstruction for the noisy Fourier intensity data with $10^4 $ photons/pixels by \textbf{(a)} HIO algorithm and \textbf{(b)} CGPR in $200$ iterations.\textbf{(c)} Phase recovery with $10^6$ photons/pixels in $200$ iterations of HIO and \textbf{(d)} CGPR algorithm.  }
\end{figure}

\section{Conclusion}
Phase retrieval from Fourier magnitude data is an active research area with important implications for a number of imaging applications in optical, coherent x-ray, and electron beam based systems. The typical iterative phase retrieval methods operate by imposing the observed Fourier magnitude constraint in the data domain and other suitable constraints (e.g. support, sparsity, etc.) in the image domain. The  imposition of constraints in the data (Fourier magnitude) domain and the image domain in these methodologies however remains somewhat disconnected. In this paper we introduced a complexity parameter that provides a measure fluctuations in the desired phase retrieval solution. It is shown that the complexity parameter can be computed directly from the Fourier magnitude data and this information can then be utilized as a guidance when applying image domain constraints such as image spasity. We call the resultant algorithm as ``Complexity Guided Phase Retrieval (CGPR)''. In the present work we have used the complexity guidance idea along with the well-known Fienup HIO algorithm. It is observed that CGPR
can provide a meaningful solution to the phase retrieval problem in significantly reduced number of phase retrieval iterations compared to the traditional HIO method. In numerical simulations we apply CGPR to Fourier intensity data (noiseless as well moderate to low Poisson noise) corresponding to phase objects and show that CGPR can also effectively address the twin stagnation issue. The ability of CGPR to provide nearly guaranteed meaningful solution with a significantly reduced number of phase retrieval iteration suggests that this methodology is valuable for designing practical applications or devices involving phase retrieval. Finally it is important to note that the complexity guidance idea can be used along with other forms of phase retrieval algorithms (e.g. more recent optimization based approaches) as well.

\end{document}